# Large-mode-area Soliton Fiber Oscillator Mode-locked with Linear Self-stabilized Interferometer

**Marvin Edelmann,**[1,2,3,*] **Malek M. Sedigheh,**[1,2] **Yi Hua,**[4] **Erwin C. Vargas,**[1,2] **Mikhail Pergament,**[1] **and Franz X. Kärtner**[1,2]

[1] *Center for Free-Electron Laser Science CFEL, Deutsches Elektronen-Synchrotron DESY, Notkestr. 85, 22607 Hamburg, Germany*
[2] *Department of Physics, Universität Hamburg, Jungiusstr. 9, 20355 Hamburg, Germany*
[3] *Cycle GmbH, Notkestr. 85, 22607 Hamburg, Germany*
[4] *Deutsches Elektronen-Synchrotron DESY, Notkestr. 85, 22607 Hamburg, Germany*
*Corresponding author: marvin.edelmann@desy.de*

Date of Submission: 10.08.2022

**In this work, we investigate an approach to scale up the output pulse energy in an all polarization-maintaining 17 MHz Yb-doped fiber oscillator via implementation of 25 µm core-diameter large-mode-area fibers. The artificial saturable absorber in form of a Kerr-type self-stabilized fiber-interferometer enables highly stable mode-locked steady-states in the soliton-like operation regime with 170 mW average output power and a total output pulse energy of ~10 nJ distributed between two output ports. An experimental parameter comparison with a reference oscillator made of 5.5 µm core-sized standard fiber-components reveals an increase of pulse energy by a factor of 36 with simultaneously reduced intensity-noise in the high frequency range > 100 kHz.**

Mode-locked fiber oscillators for the generation of stable femtosecond pulse trains are key-elements for many state-of-the-art scientific and industrial applications including synchronization and timing [1], biological imaging and spectroscopy [2], seeding of high-power lasers [3] and photonic microwave generation [4]. The output characteristics of the oscillator in conjunction with its parameter stability and resistance against environmental perturbations determine the overall system performance and robustness in industrial applications. Especially the characteristics of the saturable absorber (SA) for initiation and stabilization of the mode-locked steady-states play a crucial role as it often limits the achievable parameters with respect to optical bandwidth, intracavity power, and output pulse energy [5,6]. For more than 3 decades, artificial SA mechanisms based on the optical Kerr-effect have revealed themselves as promising candidates for next-gen fiber oscillator technology. Nowadays, fiber oscillators mode-locked with nonlinear amplifying/optical loop mirrors (NALM/NOLM) [7-9] or linear self-stabilized fiber-interferometers (LSI) [11-13] with all polarization-maintaining (PM) structures routinely achieve state-of-the-art environmental stability [10], timing jitter and intensity-noise [11]. In contrast to real SAs such as semiconductor saturable absorber mirrors (SESAM) or topological insulators, SAs based on the non-resonant optical Kerr-effect allow for large optical bandwidths and short pulse durations while being robust against optical damage and parameter degradation [12]. Besides these advantages, there are also limitations of fiber oscillators associated with the large roundtrip nonlinear phase shifts due to the strong confinement of the laser mode in the fiber segments. These nonlinear phase shifts limit the obtainable intracavity pulse energy as it initiates multi-pulse formation or breakthrough of continuous-wave lasing above a certain threshold, which ultimately deteriorates the stability of the mode-locked state [13,14]. The limitation of the pulse energy often necessitates an increased system complexity in form of additionally required amplification stages. In addition, it is further associated with additional limitations of the oscillators noise performance in terms of intensity and phase fluctuations [15,16]. Different approaches to overcome this limitation have been the subject of intensive research over the last decade. Besides techniques based on precise dispersion-management to reduce the average peak-power of the intracavity pulse in the dissipative soliton [17] or stretched-pulse regime [18], other works further revealed the possibility of periodic intracavity coherent pulse division and recombination [19,20]. In addition, the scaling of the core-size with large-mode-area (LMA) fibers in NALM/NOLM mode-locked lasers has enabled a significant increase of pulse energy due to the quadratic dependence on the core diameter [21,22].

In this work, we demonstrate the application of 25 µm core diameter LMA fibers and fiber-optic components in an LSI mode-locked fiber oscillator for the first time. The 17 MHz Yb-doped all-PM oscillator allows stable mode-locking in the soliton-like regime with a total pulse energy of 10 nJ and average power of 170 mW, distributed between two output ports. An experimental comparison with a reference oscillator operating at identical repetition rate and net-dispersion shows that the core-diameter enlargement from 5.5 µm to 25 µm results in an increase of the output pulse energy by a factor of 36 at all output ports. Further, the LSI-LMA oscillator shows an intensity-noise reduction of ~ 8 dB in the high frequency range >20 kHz compared to the reference oscillator.

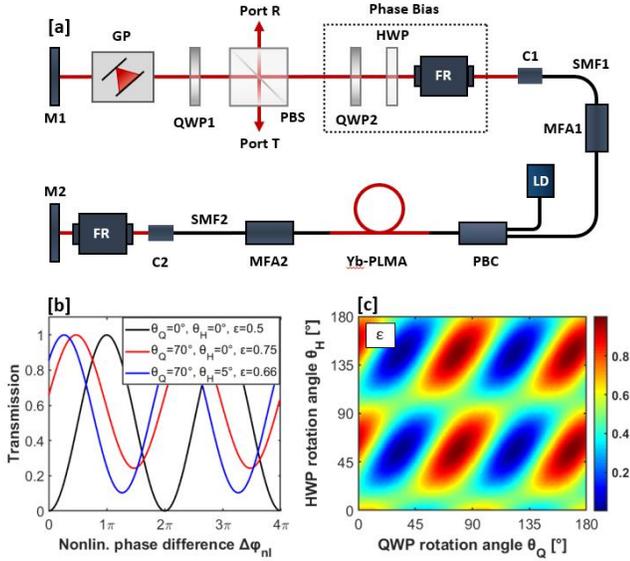

**Fig. 1:** [a]: Experimental setup of the LSI mode-locked LMA oscillator. M, mirror; GP, grating-pair; QWP, quarter-wave plate; PBS, polarization beam-splitter; FR, Faraday-rotator; C, collimator; SMF, single-mode fiber; MFA, mode-field adapter; LD, laser diode; PBC, pump-beam combiner, Yb-PLMA, ytterbium-doped polarization-maintaining large-mode-area fiber. [b]: Transmission function $T(\Delta\varphi_{nl})$ for different settings of the QWP ($\theta_Q$) and HWP ($\theta_H$) rotation angles in the phase-bias. [c]: Energy splitting-ratio $\varepsilon$ between the co-propagating orthogonal polarization modes in the PM-fiber LSI as function of $\theta_Q$ and $\theta_H$.

The experimental setup of the LSI mode-locked LMA oscillator is shown in Fig. 1[a]. The active fiber cavity includes a pump-beam combiner (PBC) to couple the pump light generated by an 18 W wavelength-stabilized multimode laser-diode at 976 nm into the cladding of a 2 m Yb-doped PM large-mode area (Yb-PLMA) fiber. The Yb-PLMA fiber (Liekki Yb1200 25/250DC-PM) has a core diameter of 25 µm, a numerical aperture of 0.065 and a peak cladding absorption of 10.6 dB/m at 976 nm. It is coiled on a 30 mm radius aluminum coil which corresponds to a theoretical propagation loss of ~2 dB and ~20 dB for the $LP_{01}$ and $LP_{11}$ mode, respectively. The thermo-conductivity of the aluminum simultaneously ensures efficient heat dissipation caused by bending-induced leakage of pump light from the cladding. As an additional filter for high-order transverse modes, the PLMA-segment is spliced to mode-field adapters (MFA1/2) at both ends, tapering the LMA fiber to 0.25 m short pieces of 5.5 µm core-diameter PM fiber (Coherent, PM980-XP). Two fiber collimators (C1/2) couple the light into free-space arms. On the side of C2, a Faraday-mirror arrangement consisting of a Faraday-rotator (FR) and the end-mirror M2 ensure the back-reflection and 90° polarization-rotation of the intracavity field. The free-space arm on the side of C1 contains a transmission grating pair (GP, 1000 lines/mm) for dispersion management, a polarization beam-splitter (PBS) and a non-reciprocal phase bias consisting of a FR, a quarter-wave plate (QWP2) and a half-wave plate (HWP1). To maintain a consistent frame of reference, the fast axis of the PM fiber-segment is aligned to the transmission axis of the PBS (subsequently referred to as x-axis) and all rotation angles are measured according to the mathematical standard definition. The repetition-rate of the laser is 17.3 MHz with a fixed net-dispersion of ~-0.121 ps$^2$, ensuring a mode-locked operation in the soliton-like regime. The mode-locking mechanism of the LSI-LMA oscillator is based on the accumulation of a differential nonlinear phase shift $\Delta\varphi_{nl}$ between the orthogonal polarization-modes in the PM-fiber segment with compensation of linear phase shifts through application of the Faraday-mirror. The artificial SA is thereby fully characterized by the sinusoidal transmission function $T(\Delta\varphi_{nl})$ that is highly tunable through the rotation angles of HWP1 ($\theta_H$) and QWP2 ($\theta_Q$) in the non-reciprocal phase-bias as indicated in Fig. 1[b]. In contrast to the well-known NALM/NOLM mode-locked lasers, the LSI does not have an inherent structural asymmetry with respect to the position of the gain fiber. Hence, the accumulation of a sufficient $\Delta\varphi_{nl}$ relies entirely on the energy splitting ratio $\varepsilon = E_x/(E_x + E_y)$ between the fast (x) and slow (y) axis of the PM fiber. The dependence of $\varepsilon$ on the phase-bias rotation angles $\theta_H$ and $\theta_Q$ is shown in Fig. 1[c]. As the phase-bias has a coupled influence on the LSI asymmetry and the $T(\Delta\varphi_{nl})$ parameters, the settings of $\theta_H$ and $\theta_Q$ have to ensure a large value of $\varepsilon$, a positive slope of $T(\Delta\varphi_{nl})$ for small signal values of $\Delta\varphi_{nl}$ (assuming a positive sign) and simultaneously a low non-saturable loss of the SA.

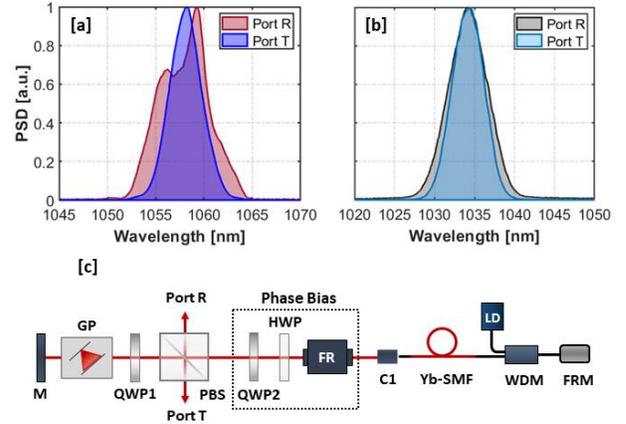

**Fig. 2:** [a]: Measured spectra at port R (red) and port T (blue) of the LSI-LMA oscillator in a soliton-like regime with a FWHM of 6 nm and ~4 nm, respectively. [b]: Measured spectra at port R (black) and port T (cyan) of the reference oscillator with a FWHM of 5 nm and 6 nm, respectively. [c]: Experimental setup of the reference oscillator. M, mirror; GP, grating-pair; QWP, quarter-wave plate; HWP, half-wave plate; FR, Faraday-rotator; C, collimator; Yb-SMF, ytterbium-doped single-mode fiber; LD, laser-diode; WDM, wavelength-division multiplexer; FRM, Faraday-rotator mirror.

In the experiment, self-starting mode-locked operation of the LSI-LMA oscillator results from a state of the SA transmission function $T(\Delta\varphi_{nl})$ with $\theta_Q = 70°$ and $\theta_Q$ in the range of 5° and 0°, corresponding to the blue and red curve in Fig. 1[b] with an energy splitting-ratio $\varepsilon$ of 0.75 and 0.66, respectively. With a net-dispersion of -0.121 ps$^2$, the laser self-starts into a multi-pulse regime at about ~6 W pump power. Reaching single-pulse operation requires a reduction of the pump power by 63 % to ~2.2 W. This well-known behavior (e.g., from NALM/NOLM oscillators) is associated with the insufficient accumulation of small-signal $\Delta\varphi_{nl}$ below a certain pump power threshold [23,24]. The continuous-wave (CW) lasing threshold is measured at 1.5 W pump power and the roundtrip cavity loss excluding the non-saturable loss from the SA is estimated to be ~65 %. A measurement of the optical spectrum in single-pulse regime at port T and port R with a center-wavelength of ~1058 nm for both cases is shown in Fig. 2[a]. The maximum obtainable output pulse energy at port T is 5.4 nJ with an average power of 92 mW, obtainable by adjusting the output coupling ratio via rotation of QWP1. Simultaneously, the pulse energy at port R reaches up to 4.3 nJ corresponding to 72 mW average power.

In order to determine the influence of the LMA-fiber configuration, the LSI-LMA oscillator performance is compared to a reference oscillator with an experimental setup as shown in Fig. 2[c]. Instead of 25 µm core-diameter LMA fibers, the fiber-segment of the reference oscillator is entirely based on standard PM fibers with a core-diameter of 5.5 µm (Nufern, PM980-XP). The 976 nm pump light is supplied by a 1 W single-mode laser diode, coupled into the 0.5 m highly Yb-doped gain fiber (CorActive, Yb401-PM) through a wavelength-division multiplexer (WDM). The free-space arm is identical to the one used for the LSI-LMA oscillator and both the repetition rate and the net-dispersion are matched to 17.4 MHz and -0.121 ps$^2$, respectively. The output pulse energy of the reference oscillator in a stable single-pulse mode-locked state corresponding to the spectra shown in Fig. 2[b] are 0.15 nJ and 0.12 nJ for port T and port R, respectively. Hence, the experimental results show that the increase of the core-size form 5.5 µm to 25 µm leads to an output energy-scaling by a factor of 36. The comparison with the reference oscillator further shows a 20 nm shift of the center-wavelength from 1038 nm to 1058 nm. This can be explained with comparable strong absorption of the signal around 1030, due to the high doping concentration in the LMA fiber that is required to compensate for the small overlap factor $\eta_p$ between the multimode pump beam and the doped fiber core. An evaluation of the doping concentration $N_{Yb} = \alpha/\sigma_{abs}\eta_p$ with the pump absorption $\alpha$ and the absorption cross-section $\sigma_{abs}$ shows an increase of $N_{Yb}$ one order of magnitude from $2.7 \ast 10^8$ µm$^{-3}$ in the Yb-SMF of the reference oscillator to $1.4 \ast 10^9$ µm$^{-3}$ in the LSI-LMA oscillator Yb-PLMA.

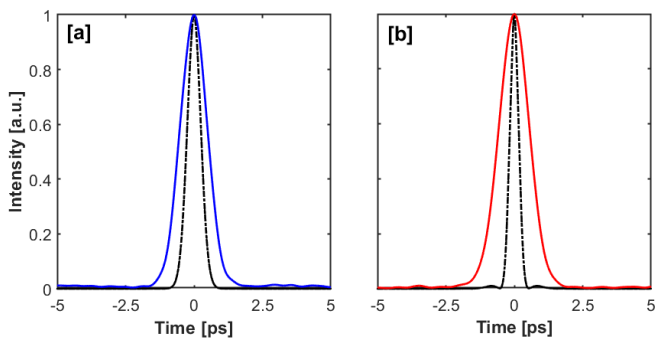

**Fig. 3:** [a]: Measured output pulse (blue) and calculated Fourier transform-limited (FTL) pulse (black) at port T of the LSI-LMA oscillator with a FWHM of 1.2 ps and 0.6 ps, respectively. [b]: Corresponding measured pulse (red) and calculated FTL pulse at port R with a FWHM of 1.3 ps and 0.36 ps, respectively.

A measurement of the autocorrelation (AC) traces at port T and port R of the LSI-LMA oscillator is shown in Fig. 3[a] and [b] together with the Fourier-transform limited (FTL) pulses retrieved from the spectra in Fig. 2. Assuming a sech$^2$ pulse shape, the pulse duration at port T in Fig. 3[a] is measured to be 1.2 ps with a calculated FTL pulse duration of ~0.6 ps. The output pulse at Port T is negatively chirped due to the propagation of an average soliton according to the master equation [25] in conjunction with the structure of the oscillator shown in Fig. 1[a]. For the same reason, the measured output pulse at port R is positively chirped in Fig. 3[b] with a FWHM of 1.3 ps compared to the corresponding FTL with a duration of 0.36 ps.

To evaluate and classify the stability and noise performance of the LSI-LMA oscillator, the frequency-resolved intensity-fluctuations of the reflected and transmitted output ports are investigated. As a first step of the measurement, the pulse train at the respective port is detected with a fast and low-noise photo-detector (Thorlabs DET08CFC) and the 8$^{th}$ harmonic of the received RF-signal at 139.2 MHz is filtered with a tunable bandpass-filter (BPF). Subsequently, the RF-signal is amplified with a 10 dB low-noise trans-impedance amplifier (MiniCircuits ZX60-33LN-S+) to an RF-power of -6.5 dBm, corresponding to a consistent shot-noise level at -142.3 dBc. The amplitude-modulation (AM) function of a signal-source analyzer (SSA, Keysight E552B) is then used to measure the single side-band frequency-resolved AM-noise spectral density. The results of the AM-noise measurements for the LSI-LMA and the reference oscillator at port T and port R are shown in Fig. 4[a] and [b], respectively. The measured traces include the frequency-range from 10 Hz to 5 MHz, limited by the bandwidth of the BPF.

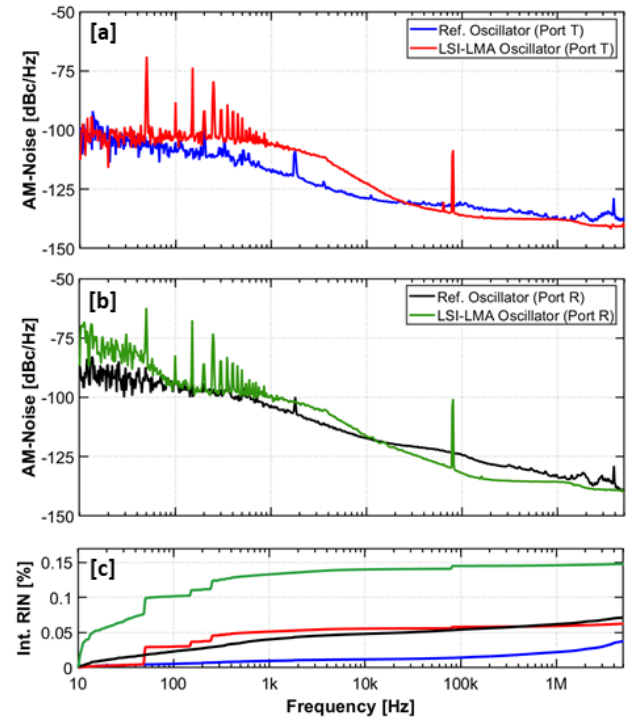

**Fig. 4:** [a]: Frequency-resolved amplitude-noise measured at port T of the reference oscillator (blue) and LSI-LMA oscillator (red). [b]: AM-noise spectra measured at port R of the LSI-LMA (green) and reference oscillator (black). [c]: Corresponding relative intensity-noise (RIN) integrated from 10 Hz to 5 MHz of 0.063% (LSI-LMA) and 0.038% (reference) at port T and 0.148% (LSI-LMA) and 0.071% (reference) at port R, respectively.

In comparison with the reference oscillator, the magnitude of the AM-noise measured at port T shows a high similarity in the frequency-range <50 Hz due to the identical environment of both laser systems in the laboratory. In the range from 50 Hz to ~20 kHz, the noise of the LSI-LMA oscillator is significantly higher with a difference of up to 12 dB. A defining characteristic of the LMA oscillators AM-noise spectra in this frequency range is a comb of resonant peaks at the utility frequency of 50 Hz and its higher harmonics. Hence, there is a strong indication that the increased AM-noise is not caused by intrinsic optical properties of the LSI-LMA laser nor the characteristics of the multimode LD, but instead by the influence of the power supply with insufficient electronic filtering/shielding. As a consequence, the RIN integrated over the full measurement bandwidth reaches 0.063% for the LSI-LMA and only 0.038% for the reference oscillator as shown in Fig.4 [c]. In the frequency range >20 kHz, associated with the fast optical dynamics in the cavity, the LSI-LMA port T output shows an improved AM-noise performance with a decreased magnitude of the fluctuations by up to 5 dB and an integrated RIN (20 kHz - 5MHz integration range) of 0.027% compared to 0.033% from the reference oscillator. An almost identical behavior can be observed in the AM-noise spectra and the RIN measured at port R as shown in Fig.4 [b] and [c], respectively. The AM-noise of the LSI-LMA output pulse train at port R has an increased magnitude in the frequency range from 50 Hz 20 kHz compared to the reference oscillator, while it performs significantly better for frequencies >20 kHz with an improvement of up to 8 dB. With an integration over the complete measurement bandwidth, the RIN at port R reaches 0.148% in case of the LSI-LMA oscillator and 0.071% for the reference oscillator. In the frequency-range >20 kHz, the RIN (20 kHz-5MHz) of the LSI-LMA oscillator is lower with a value of 0.045% compared to 0.050% for the reference oscillator. A comparison of the fluctuations measured at port T and port R shows a noise difference for both laser systems. In both cases, the AM-noise measured at port T is significantly lower than at port R, leading to a difference in integrated RIN between the output ports of ~3.7 dB in the case of the LSI-LMA oscillator and ~2.7 dB for the reference oscillator. This noise difference is characteristic for fiber oscillators mode-locked with nonlinear fiber interferometers and can be explained with the dynamic response of the SA transmission $T(\Delta\varphi_{nl})$ to the intensity-fluctuations of the intracavity pulse and thus in agreement with our study summarized in Ref. [26].

In conclusion, we demonstrated an investigated an Yb-doped all-PM fiber oscillator mode-locked with a self-stabilized Sagnac-interferometer that uses a 25 μm core LMA fiber to avoid excessive nonlinear phase-shifts while scaling up the energy. Stable and self-starting mode-locked steady-states in the soliton-like regime are demonstrated with a total pulse energy of ~10 nJ and 170 mW average power distributed between the two output ports. An experimental comparison of the LSI-LMA oscillator with a reference oscillator constructed with 5.5 μm standard core-diameter fiber components shows that the implementation of 25 μm core LMA fibers allow an energy scaling by a factor of 36. In comparison to the reference oscillator, the AM-noise spectral density of the LMA oscillator further shows a significant improvement for high offset-frequencies > 20 kHz by up to 8 dB. The results of this work demonstrate the possibility to implement LMA fibers in fiber oscillators mode-locked with LSI for highly-efficient energy-scaling while preserving and even improving the stability and noise-performance of the laser system. The successful demonstration of the LSI-LMA oscillator configuration is another important step in the ongoing search for high-power, environmentally stable and ultra-low noise fiber oscillators for the generation of femtosecond pulse trains.